\documentclass[aps,prl,twocolumn,superscriptaddress,accepted=2018-08-06]{quantumarticle}
\pdfoutput=1

\usepackage{amsmath,amsfonts,amssymb,graphicx,xcolor,natbib}
\usepackage{tikz}
\usetikzlibrary{arrows,fit,positioning,automata,fadings,arrows.meta}
\usetikzlibrary{decorations,decorations.pathmorphing,decorations.pathreplacing,shapes.arrows,shapes.geometric}

\def\spvecA#1;{\if;#1;\else #1\cr \expandafter \spvecA \fi}

\DeclareMathOperator\atanh{atanh}

\DeclareGraphicsExtensions{.eps, .pdf,.png,.jpg}

\begin{document}

\title{Measuring acceleration using the Purcell effect}

\author{Kacper Ko\.{z}do\'{n}}
\affiliation{Institute of Theoretical Physics, Faculty of Physics, University of Warsaw, Pasteura 5, 02-093 Warsaw, Poland}
\author{Ian T. Durham}
\email[]{idurham@anselm.edu}
\affiliation{Department of Physics, Saint Anselm College, Manchester, NH 03102, USA}
\author{Andrzej Dragan}
\affiliation{Institute of Theoretical Physics, Faculty of Physics, University of Warsaw, Pasteura 5, 02-093 Warsaw, Poland}
\date{\today}

\begin{abstract}
We show that a two-level atom resonantly coupled to one of the modes of a cavity field can be used as a sensitive tool to measure the proper acceleration of a combined atom-cavity system. To achieve it we investigate the relation between the transition probability of a two-level atom placed within an ideal cavity and study how it is affected by the acceleration of the whole. We indicate how to choose the position of the atom as well as its characteristic frequency in order to maximize the sensitivity to acceleration.
\end{abstract}
 
\maketitle

The spontaneous emission of a photon from an atom is a property of the atom-vacuum system, rather than of the atom itself. The irreversibility of such systems arises from the fact that an infinite number of vacuum states is typically available to the radiated photon. Modifications to the vacuum states can thus be used to either inhibit or enhance the spontaneous emission. One such modification involves placing an excited atom between mirrors in an optical cavity. In the weak coupling limit between a two-level atom and a single-mode cavity, the resulting change to the transition rate of the atom is known as the Purcell effect~\cite{Purcell:1946aa,Kleppner:1981aa}.

Additional modifications to the vacuum states occur as a result of acceleration. The Unruh effect, for instance, implies that an observer in an accelerated reference frame should detect the presence of a background thermal bath of photons where an inertial observer would detect none~\cite{Fulling:1973aa,Davies:1975aa,Unruh:1976aa}. Thus, for an accelerated atom, there are no longer an infinite number of vacuum states available to a radiated photon which makes the atom theoretically capable of detecting the non-inertial background bath. Unfortunately, this effect is very weak and has not yet been detected, despite continuing efforts.

In this work we investigate another effect of acceleration on the mode structure of the vacuum state that can significantly modify the atom-field interaction. We consider an ideal optical cavity in uniformly accelerated motion with an atom placed inside that is co-accelerating with the cavity. We show how to position the atom and tune its interaction with the cavity field in order to maximize the sensitivity of the transition rate of such a system to the acceleration. 

Specifically we consider an Unruh-DeWitt detector which is a simplified model of an atom moving along an arbitrary classical trajectory~\cite{Unruh:1976aa,DeWitt:1975aa}. Accelerated Unruh-DeWitt detectors in stationary cavities have been analyzed, but the observed effects can be attributed to the relative acceleration that exists between the detectors and the cavity~\cite{Scully:2003aa,Belyanin:2006aa,Dragan:2010aa,Abdolrahimi:2014aa,Ahmadzadegan:2014ab,Ahmadzadegan:2014aa}. Solutions have also been found for scalar fields in cavities with a time-varying size, though these results only apply to the scalar field modes of the cavities themselves and lack any coupling to detectors~\cite{Koehn:2012aa,Koehn:2013aa,Bruschi:2013aa,Bialynicki-Birula:2013aa}.

It is worth asking, then, what the effect of the accelerated motion of the combined atom-cavity system is on the dynamics of the quantum state of this system. In the weak coupling limit, the atomic transition is irreversible. In this limit the presence of the cavity necessarily modifies the vacuum states (via the Purcell effect), and thus any changes to the dynamics of the system will manifest as changes to the Purcell effect if the entire system, consisting of detector and cavity, is accelerated as one. In the strong coupling limit, the atomic transition is reversible meaning an emitted photon can be reabsorbed by the atom before escaping the cavity. In this article we confine ourselves to discussion of the weak coupling limit.

We begin by considering the probability for a detector to transition between an excited state $|e\rangle$ and a ground state $|g\rangle$. Such a transition necessarily involves the emission of a photon whose probability amplitude is modeled by the Klein-Gordon equation. We consider both massive and massless 1+1 dimensional real scalar field models ($\hbar = c = 1$):
\begin{equation}\label{eqn: kge}
(\Box + m^2)\hat{\phi}(x, t) = 0
\end{equation}
with the canonical scalar product
\begin{equation}\label{eqn: sca}
\left(\hat{\psi}_1, \hat{\psi}_2\right) = -i\int\mbox{d}x\left(\hat{\psi}^\ast_1\overleftrightarrow{\partial}_\tau\hat{\psi}_2\right).
\end{equation}
When considered in the interior of a stationary cavity of proper-length $L$, the walls of the cavity introduce Dirichlet boundary conditions and the solution becomes
\begin{equation}
\label{decompfield}
\hat{\phi}(t, x) = \displaystyle\sum_{k = 1}^\infty F_k(x)\left(e^{-i\omega_kt}\hat{a}_k + e^{i\omega_kt}\hat{a}_k^\dagger\right),
\end{equation}
where the $F_k(x)$ describe eigenmodes of the cavity with the corresponding quantized frequencies $\omega_k$. For a cavity at rest, the spatial part of the solution is given by:
\begin{equation}\label{eqn: inrtl}
\left\{\begin{array}{l}
F_k(x) = \frac{1}{\sqrt{\omega_k L}}\sin\left(k\pi\frac{x + \frac{L}{2}}{L}\right)\\
\\
\omega_k=\sqrt{\left(\frac{k\pi}{L}\right)^2 + m^2}\mbox{.}
\end{array}\right.
\end{equation}

Let us now consider a single Unruh-DeWitt detector governed by the interaction Hamiltonian:
\begin{equation}\label{eqn: hint}
\hat{H}_I(\tau)\propto\epsilon(\tau)\hat{\phi}(x(\tau))\left(e^{-i\omega\tau}\hat{d} + e^{i\omega\tau}\hat{d}^\dagger\right)\mbox{,}
\end{equation}
with $\tau$ being the proper time along the detector's trajectory $x(\tau)$ and with $\epsilon(\tau)$ representing the coupling strength. The annihilation operator and characteristic frequency of the detector are $\hat{d}$ and $\omega$ respectively. We can then position the detector in the center of the cavity such that it always lies on the cavity's reference trajectory and is thus always at rest with respect to it. Initially, the detector is in its excited state $|e\rangle$ and we calculate the amplitude of a transition to its ground state $|g\rangle$ in the lowest order in $\epsilon$. We will assume that the cavity is initially in the vacuum state of all the modes in its co-moving reference frame. Due to the interaction between the detector and the cavity field, the transition from $|e\rangle$ to $|g\rangle$ has a non-zero probability, which we will minimize by tuning the detector frequency to match the frequency of the second eigenmode of the cavity. In this case, the atom couples most strongly to the second cavity mode which vanishes at the position of the detector when the cavity is at rest. In such a situation, only the off-resonant interaction with the remaining modes contributes to the transition probability and even then it will become negligible for sufficiently long interaction times. As a result, such an atom in the middle of the cavity will remain in its excited state provided the cavity remains at rest.

To show this we compute the first-order probability amplitude for the atom to make a transition to the ground state with the field going from the initial vacuum state to some arbitrary final state $|\psi\rangle$:
\begin{equation}\label{eqn: amp}
\mathcal{A}_\psi=-i\int\limits_{-\infty}^{\infty}\mbox{d}\tau\left<g\right|\left<\psi\right|\hat{H}_I(\tau)\left|e\right>\left|0\right>.
\end{equation}
The total probability for the detector ending up in the ground state is given by:
\begin{equation}\label{eqn: prob}
\mathcal{P}=\displaystyle\sum_\psi\left|\mathcal{A}_\psi\right|^2.
\end{equation}
Using the decomposition \eqref{decompfield} one finds that the probability of the decay is given by the following sum of integral expressions:
\begin{equation}
\begin{split}\label{eqn: probsum}
\mathcal{P} = \displaystyle\sum_{k = 1}^\infty\left|\int\mbox{d}\tau\epsilon(\tau)F_k\left(x(\tau)\right)e^{-i(\omega\tau - \omega_kt(\tau))}\right|^2 \mbox{.}
\end{split}
\end{equation}
For a resting detector switched on for a time $\tau$ this yields:
\begin{widetext}
\begin{equation}\label{eqn: prest}
\mathcal{P}_{\textrm{rest}} =  \left|\tau\frac{\epsilon}{\sqrt{\omega_2 L}}\sin\left(2\pi\frac{x_0 + \frac{L}{2}}{L}\right)\right|^2 +  \displaystyle\sum_{k\neq 2}\left|\frac{\epsilon}{\omega_k - \omega_2}\frac{1}{\sqrt{\omega_k L}}\sin\left(k\pi\frac{x_0 + \frac{L}{2}}{L}\right)\left(e^{-i(\omega_2 - \omega_k)\tau}-1\right)\right|^2\mbox{,}
\end{equation}
\end{widetext}
where $x_0 = 0$ corresponds to the position of the detector inside the cavity. As we can see, the first (resonant) term vanishes and, for finite interaction times, the remaining off-resonant, oscillating terms fall off quickly as $|\omega_k-\omega_2|$ increases, making the total transition probability approximately zero.

Let us now consider the situation in which the detector and the cavity comove with a uniform proper acceleration along the cavity's length. In this case the cavity will relativistically contract~\cite{Dewan:1959aa,Bell:1976aa} in which case the cavity's second eigenmode will not vanish at the position of the detector thus making the transition to the ground state possible. As a consequence, a transition from the excited state to the ground state can serve as an indication that the system is accelerating. As such, the system consisting of the detector and cavity together functions as an accelerometer.

In order to find stationary solutions for a uniformly accelerated cavity, we introduce Rindler coordinates $(\tau,\chi)$ for the accelerated frame,
\begin{equation}
\left\{\begin{array}{l}
\tau = \frac{1}{a}\atanh\left(\frac{t}{x}\right) \\
\\
\chi = \sqrt{x^2 - t^2},
\end{array}\right.
\end{equation}
where $a$ is a parameter corresponding to the proper acceleration of an arbitrarily chosen accelerated reference trajectory. As mentioned above, we will choose to have this reference trajectory lie in the very center of the cavity, so that when the length $L$ of the cavity is much shorter than $a^{-1}$, the parameter $a$ can be treated as the proper-acceleration of the entire cavity. In this case $\tau$, being the proper time along the reference trajectory, then becomes the proper time of the entire cavity. The eigenmodes of the accelerated cavity are then the solutions of 
\begin{equation}
\left(\displaystyle\frac{1}{a^2\chi^2}\partial^2_\tau - \partial^2_\chi - \frac{1}{\chi}\partial_\chi + m^2\right)\hat{\phi} = 0
\end{equation}
which is the massive Klein-Gordon equation (\ref{eqn: kge}) in Rindler coordinates. General solutions to this equation are of the form given in~\eqref{decompfield} where the eigenmodes can now be expressed in terms of modified Bessel functions of the first kind with purely imaginary order $I_{\pm i\nu}\left(m\chi\right)$~\cite{Dragan:2010aa,Dunster:1990aa}:
\begin{align}
F_{\Omega_k}(\chi) & = N_k\left(I_{i\frac{\Omega_k}{a}}\left(m\chi\right)I_{-i\frac{\Omega_k}{a}}\left(m\chi_2\right) \right. \nonumber \\ & \left.- I_{-i\frac{\Omega_k}{a}}\left(m\chi\right)I_{i\frac{\Omega_k}{a}}\left(m\chi_2\right)\right), \label{eqn: solacc}
\end{align}
where $\Omega_k$ is the corresponding eigenfrequency and $N_k$ is a normalization constant. We of course choose the boundary conditions such that the eigenmodes vanish at the positions of the cavity mirrors, $\chi_1 = \frac{1}{a} - \frac{L}{2}$, $\chi_2 = \frac{1}{a} + \frac{L}{2}$, which necessarily quantizes $\Omega_k$. 

\begin{figure}
\includegraphics[angle=0, width=0.46\textwidth]{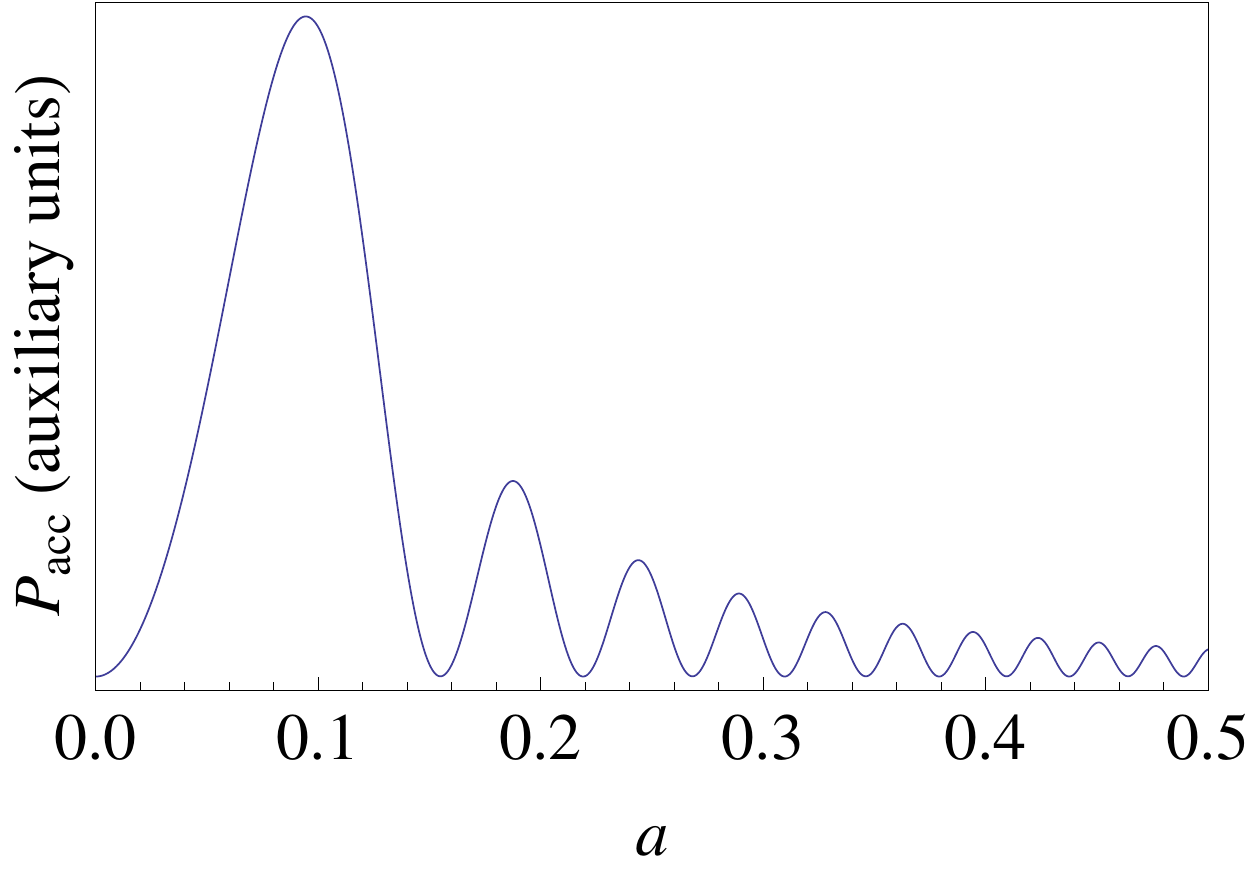}
\caption{Decay probability as a function of the proper acceleration $a$ of the cavity for arbitrary $m$ and $\epsilon$.}\label{fig: clo}
\end{figure}

When the cavity and detector are subjected to the same uniform acceleration, (\ref{eqn: probsum}) becomes:
\begin{widetext}
\begin{equation}\label{eqn: pacc}
\begin{split}
\mathcal{P}_{\textrm{acc}}  = \displaystyle\sum_{k} \left|\frac{\epsilon}{\Omega_k - \omega_2}N_k\left(I_{i\frac{\Omega_k}{a}}\left(m\frac{1}{a}\right)I_{-i\frac{\Omega_k}{a}}\left(m\chi_2\right) - I_{-i\frac{\Omega_k}{a}}\left(m\frac{1}{a}\right)I_{i\frac{\Omega_k}{a}}\left(m\chi_2\right)\right)\left(e^{i(\Omega_k - \omega_2)\tau} - 1\right)\right|^2.
\end{split}
\end{equation}
\end{widetext}
We note that, due to the semi-arbitariness of the coupling constant $\epsilon$ (we are only restricting ourselves to the weak coupling regime), the normalization constant is also arbitrary. As such we are working in arbitrary units. For an equally arbitrary, non-zero mass, we have plotted (\ref{eqn: pacc}) as a function of the proper acceleration of the cavity for the second field mode in Fig.~\ref{fig: clo}.
We see that the probability rises rapidly for small accelerations before beginning to oscillate. At the same time, the local maxima decrease with $a$.

\begin{figure}
\includegraphics[angle=0, width=0.46\textwidth]{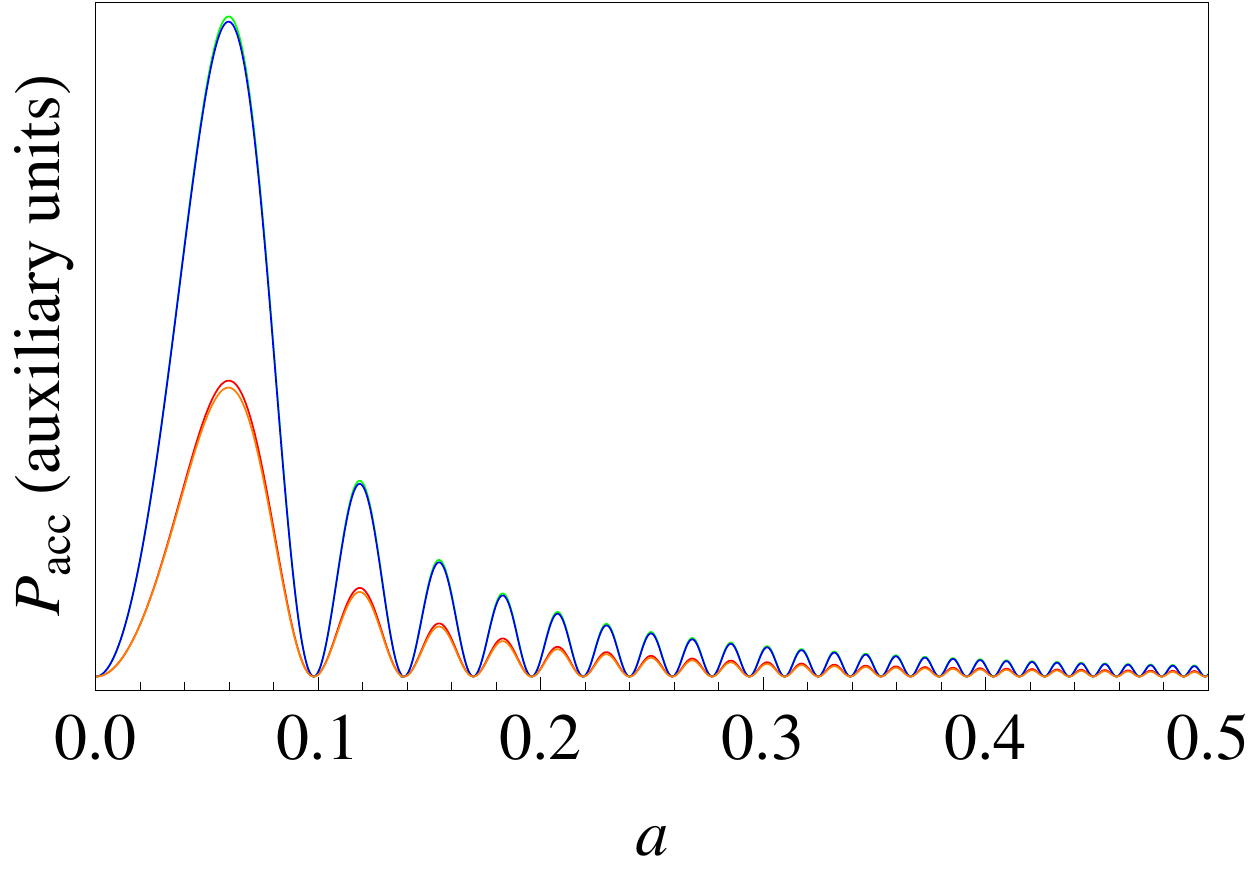}
\caption{Dependence of the decay probability on acceleration for the detector coupled to the fifth field mode. The probability for each node is given from the rearmost node forward by the red, green, blue, and orange lines respectively.}\label{fig: 395}
\end{figure}

If we were to choose to couple the detector to a higher field mode, we would have multiple nodes at which we could choose to place the detector. For the resonant detector's frequency $\omega_n$ this corresponds to placing the detector at positions $\frac{1}{a} - \frac{L}{2} + \frac{kL}{n}\mbox{, } k = 1\ldots n -1$ at which there would be no decay in a resting cavity. We find that the effect is not uniform across the nodes as shown in Figs.~\ref{fig: 415}-\ref{fig: 395}. 
\begin{figure}
\begin{center}
\begin{tikzpicture}
\node at (0,0) {\includegraphics[angle=0, width=0.46\textwidth]{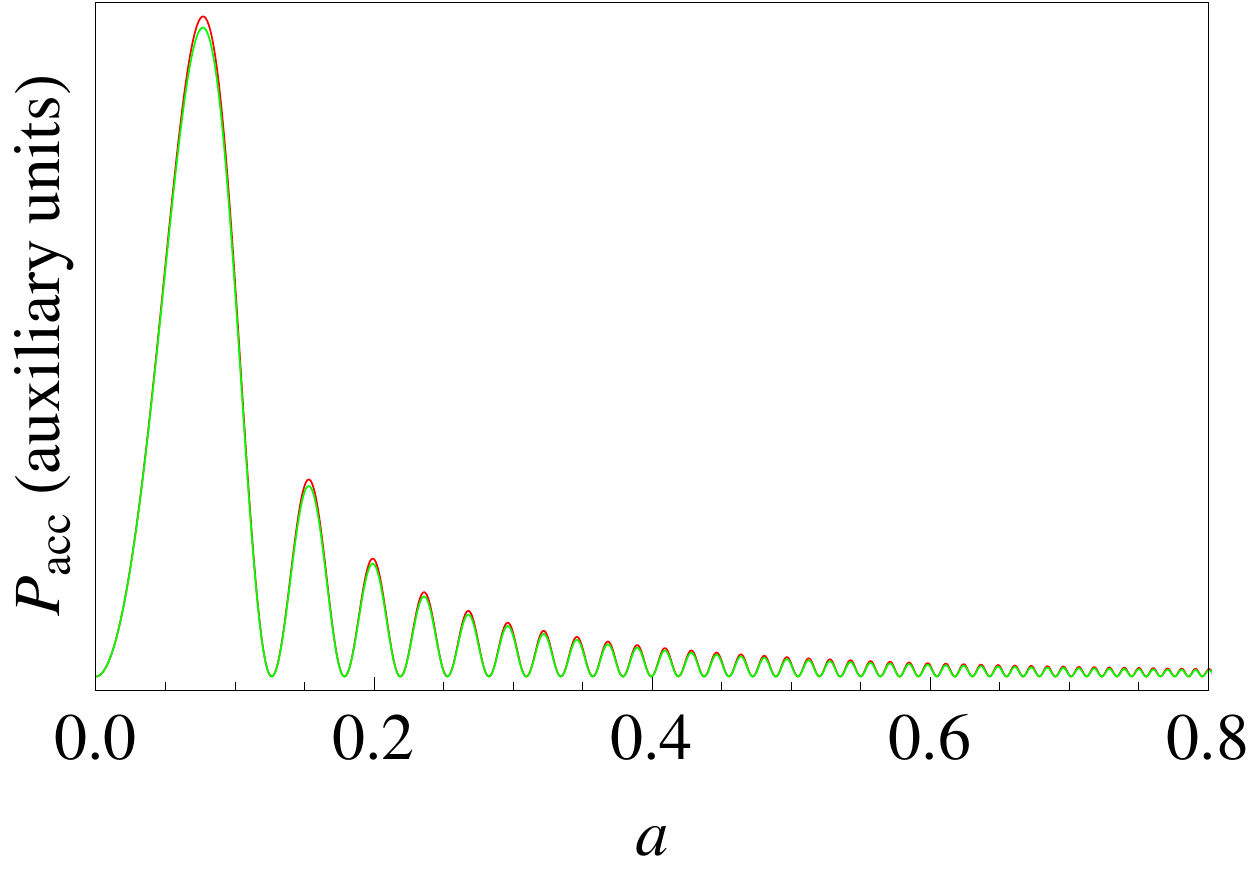}};
\node at (1.25,0.85) {
   \begin{tikzpicture}[scale=0.75]
   \draw (-0.5,-0.5) rectangle (6,3.15);
   \fill[white!50!gray] (0,0) rectangle (0.25,2);
   \node[rotate=90] at (-0.25,1) {\footnotesize{rear mirror}};
   \fill[white!50!gray] (5.25,0) rectangle (5.5,2);
   \node[rotate=90] at (5.75,1) {\footnotesize{front mirror}};
   \fill[thick,blue,opacity=0.25] (0.25,1) sin (1.083,2) cos (1.917,1) sin (2.75,0) cos (3.583,1) sin (4.417,2) cos (5.25,1);
   \fill[thick,blue,opacity=0.25] (0.25,1) sin (1.083,0) cos (1.917,1) sin (2.75,2) cos (3.583,1) sin (4.417,0) cos (5.25,1);
   \draw[thick] (0.25,1) sin (1.083,2) cos (1.917,1) sin (2.75,0) cos (3.583,1) sin (4.417,2) cos (5.25,1);
   \draw[thick] (0.25,1) sin (1.083,0) cos (1.917,1) sin (2.75,2) cos (3.583,1) sin (4.417,0) cos (5.25,1);
   \node[red] at (1.917,1) {$\bullet$};
   \node[green] at (3.583,1) {$\bullet$};
   \draw[ultra thick,-latex] (1.083,2.5) -- (4.417,2.5);
   \node at (2.75,2.85) {\footnotesize{direction of acceleration}};
   \draw[latex-latex] (1.083,2) -- (1.083,0);
   \node[left] at (1.083,1) {$\mathcal{A}$};
   \end{tikzpicture}
   };
\end{tikzpicture}
\end{center}
\caption{\label{fig: 415} The inset shows the probability amplitude for the third field mode as a function of location in a simple 1D cavity \textit{at rest}. It also shows the direction that this initially resting cavity is accelerated. The plot shows the probability amplitude as a function of $a$ for the \textit{accelerated} cavity. The green line corresponds to a detector positioned at the green node in the inset while the red line corresponds to a detector positioned at the red node in the inset.}
\end{figure}
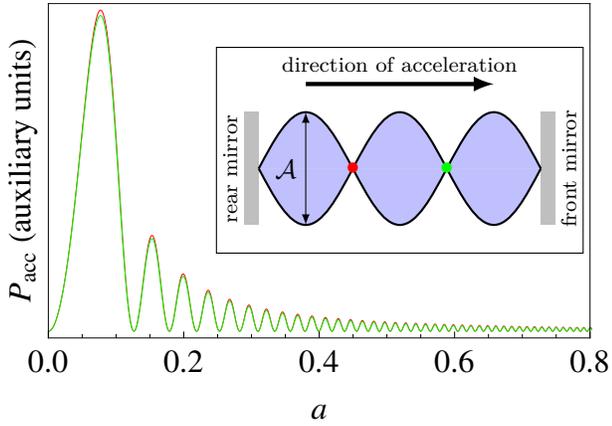
\begin{figure}
\includegraphics[angle=0, width=0.46\textwidth]{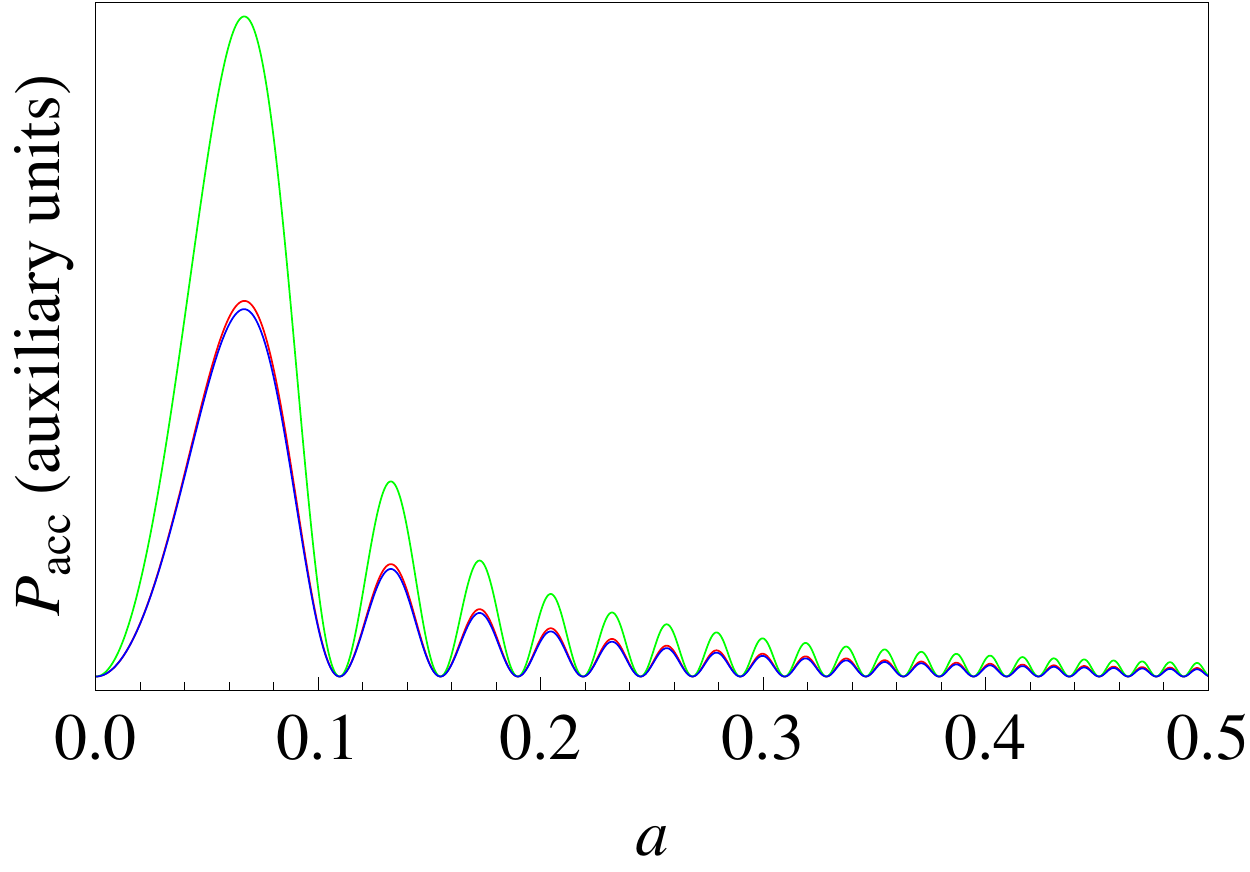}
\caption{Dependence of the decay probability on acceleration for the detector coupled to the fourth field mode. The probability for each node is given from the rearmost node forward by the red, green, and blue lines respectively.}\label{fig: 405}
\end{figure}
The effect is greatest in the node in the middle of the cavity, and if there is no node in the middle, then the node closest to the middle in the direction of acceleration (see inset, Fig.~\ref{fig: 415}).

\begin{figure}
\includegraphics[angle=0, width=0.46\textwidth]{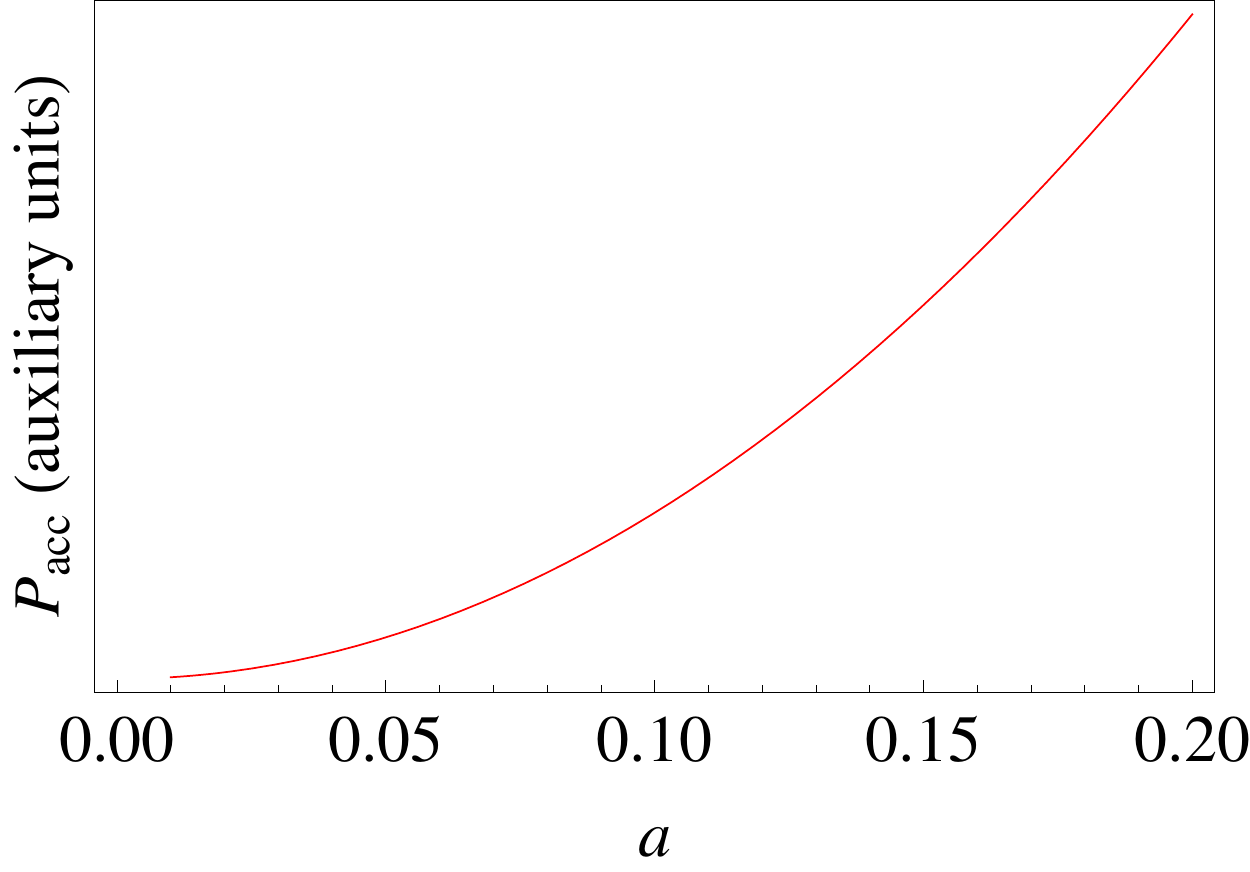}
\caption{Decay probability as a function of the proper acceleration $a$ of the cavity for a massless scalar field.}\label{fig: msl}
\end{figure}

The situation is quite different for massless fields. In that case, the spatial part of the solution to the Klein-Gordon equation takes the form:
\begin{equation}\label{eqn: msls}
\left\{\begin{array}{l}
F_{\Omega_k\mbox{, } m = 0}(\chi) = \frac{1}{\sqrt{k\pi}}\sin\left(\Omega_k\left(\xi - \xi_l\right)\right) \\
\\
\Omega_k = \frac{k\pi}{\xi_2 - \xi_1}\\
\\
\xi = \frac{1}{a}\log(a\chi)\\
\\
\xi_{i} =  \frac{1}{a}\log(a\chi_{i}), \>i=1, 2
\end{array}\right.
\end{equation}
which is just (\ref{eqn: inrtl}) with $x\to \xi(a)$ and $\omega_k \to \Omega_k$. The decay rate is then 
\begin{equation}
\left\{\begin{array}{l}
\begin{split}
\mathcal{P}_{acc}(a)\stackrel{\Omega_k\neq\omega_2}{=}&\displaystyle\sum_k \frac{1}{k\pi}\sin^2\left(k\pi\frac{-\log (1 - \frac{a L}{2})}{L^\prime a}\right)\\ & \times\frac{1}{(\omega_2 - \Omega_k)^2}\left(e^{-i(\omega_2 - \Omega_k)\tau} - 1\right)^2\end{split}\\
\\
L^\prime = \frac{1}{a}\log(\frac{1+\frac{aL}{2}}{1-\frac{aL}{2}})\mbox{.}
\end{array}\right.
\end{equation} 
 a plot of which is shown in Fig.~\ref{fig: msl}.
Notice that there is no oscillatory behavior for the massless scalar field in the considered range, even though we observe such a behavior in massive case. In addition, the decay probability increases steadily with increasing $a$ whereas for the massive field, the local maxima for the decay probability \textit{decrease} with increasing $a$, eventually dying out entirely.

In our simplified detector model we have ignored the relativistic effects of acceleration on the atom itself. Such effects are described in~\cite{Marino:2014aa}. In other words, in our model we are only considering changes to the cavity itself whereas accelerated atoms will experience changes to their energy levels independent of the presence of the cavity. In addition, realistic models would be three-dimensional. For a single-mode cavity, the transition rate can be computed from the density of states and Fermi's Golden rule and is a function of several factors that can be affected by relativistic acceleration. Thus additional work would need to be done to fully understand the relativistic effects on a more realistic model.

That said, experimental realization of such a model might be difficult. A condensed matter analog does potentially exist, though. Work has been done on frustrated spontaneous emission in metamaterial photonic band gaps. For example, work by Ginzberg, et. al. has demonstrated an analog Purcell effect in a nonlocal metamaterial that is based on a plasmonic nanorod assembly~\cite{Ginzburg:aa}. Observing the changes to such a metamaterial under relativistic acceleration would potentially serve as an experimental test bed for our model, though additional work would need to be done to ensure that the analogy holds.

\begin{acknowledgements}
We thank Jorma Louko and Keith Schwab for useful discussions and Mario Serna for alerting us to the work on metamaterials. We also thank an anonymous reviewer for helpful comments. ITD acknowledges FQXi and the Silicon Valley Community Foundation for funding. AD acknowledges support from the National Science Centre, Sonata BIS Grant No. 2012/07/E/ST2/01402.
\end{acknowledgements}

\bibliography{cavity}
\end{document}